\documentclass[12pt]{article}
\usepackage{amsmath,amssymb}
\usepackage{graphicx}

\renewcommand{\theequation}{\arabic{equation}}
\renewcommand{\thesection}{\arabic{section}}

\renewcommand{\thefootnote}{\fnsymbol{footnote}}
\newcommand{\bea}{\begin{eqnarray}}
\newcommand{\ena}{\end{eqnarray}}
\newcommand{\vs}[1]{\vspace{#1 mm}}

\renewcommand{\a}{\alpha}
\renewcommand{\b}{\beta}

\newcommand{\n}{\nu}

\renewcommand{\t}{\tau}

\newcommand{\PL}[1]{Phys.\ Lett.\ {\bf #1}}

\newcommand{\PR}[1]{Phys.\ Rev.\ {\bf #1}}

\newcommand{\PTP}[1]{Prog.\ Theor.\ Phys.\ {\bf #1}}

\newcommand{\NP}[1]{Nucl.\ Phys.\ {\bf #1}}

\begin{document}
\noindent

\begin{titlepage}
\setcounter{page}{0}
\begin{flushright}
September, 2002\\
hep-ph/0210051\\
OU-HET 420\\
\end{flushright}
\vs{2}
\begin{center}
{\Large{\bf Effects of charged Higgs boson and QCD corrections in $\bar B \to D \tau \bar\nu_{\tau}$}}
\footnote{Talk given by T. Miura at 
3rd Workshop on Higher Luminosity B Factory, 
 August 6-7, 2002, Shonan, Japan}\\
\vs{8}
{\large
Tsutomu MIKI, 
Takahiro MIURA\footnote{e-mail address:
miura@het.phys.sci.osaka-u.ac.jp}
and Minoru TANAKA\footnote{e-mail address:
tanaka@phys.sci.osaka-u.ac.jp}\\
\vs{2}
{\sl Department of Physics,
Osaka University \\ Toyonaka, Osaka 560-0043, Japan}
}
\end{center}
\vs{8}
\begin{abstract}
We study effects of charged Higgs boson exchange in 
$\bar B \to D \tau \bar\nu_{\tau}$. 
The Yukawa couplings of Model II of two-Higgs-doublet model, 
which has the same Yukawa couplings as MSSM, is considered.
We evaluate the decay rate including next-to-leading QCD corrections
and estimate uncertainties in the theoretical calculation.
Our analysis will contribute to probe an extended Higgs sector at B factory 
experiments.
\end{abstract}

\end{titlepage}
\newpage
\vskip 2cm
\renewcommand{\thefootnote}{\arabic{footnote}}
\setcounter{footnote}{0}

\section{Introduction}

The minimal supersymmetric standard model (MSSM) \cite{MSSM} 
is one of the most attractive models beyond the standard model (SM). 
In the MSSM, 
two Higgs doublets are introduced in order to 
cancel the anomaly and to give the fermion masses.
The introduction of the second Higgs doublet inevitably means that 
a charged Higgs boson is in the physical spectra.
So, it is quite important
to study effects of the charged Higgs boson.

Here, we study effects of the charged Higgs boson 
on the exclusive 
semi-tauonic $B$ decay, $\bar B \to D \tau \bar\nu_{\tau}$,
in the MSSM. 
In our previous works \cite{MT,Tanaka}, 
we calculated the decay rate of $\bar B \to D \tau \bar\nu_{\tau}$ 
including the effect of charged Higgs boson exchange 
in the leading logarithmic approximation and the heavy quark limit.
In this work, we show the decay rate 
with QCD corrections up to the next-to-leading order (NLO).
The NLO corrections are necessary to estimate theoretical uncertainties
coming from short-distance calculations 
in the ratio of the decay rates (see below).
In addition, these corrections may cause dominant uncertainties 
for the $q^2$ distribution \cite{Soni}
and the $\tau$ polarization \cite{Tanaka}.

In a two-Higgs-doublet model, 
the couplings of charged Higgs bosons to quarks and leptons are given by 
\bea
   {\cal L}_H&=&(2\sqrt{2}G_F)^{1/2}
    \bigl[X\overline{u}_L V_{KM}M_d d_R+
          Y\overline{u}_R M_u V_{KM}d_L
          +Z\overline{\nu}_L M_l l_R\bigr]\;H^+
            \nonumber\\
&&+\mbox{\large h.c.}\;,\label{Lag}
\ena
where $M_u$, $M_d$ and $M_l$ are diagonal quark and lepton mass matrices,
and $V_{KM}$ is Kobayashi-Maskawa matrix \cite{CKM}.
In the MSSM, we obtain 
\bea
   X=Z=\tan\beta\;,\;\;Y=\cot\beta\;,\label{XYZ}
\ena
where $\tan\beta=v_2/v_1$ is the ratio of the vacuum expectation 
values of the Higgs bosons. Since the Yukawa couplings of the MSSM are 
the same as those of the so-called Model II of 
two-Higgs-doublet models \cite{HHG}, 
the above equations and the following results apply 
to the latter as well.\footnote{
It is known that SUSY loop effects in Eq.(\ref{XYZ}) are significant
for large $\tan\b$ \cite{Wagner}. The dominant effect in the 
$b\to c\t\bar\n_{\t}$ decay \cite{Sola} comes from the SUSY-QCD correction
to the bottom quark mass \cite{SUSY-QCD}. 
Once this correction is taken into account, the Yukawa couplings of the MSSM 
is no longer the same as Model II of two-Higgs-doublet models.
This effect cannot be ignored in order to study the MSSM Higgs sector.
However, we omit it in this work because our aim is to clarify 
low-energy QCD corrections and uncertainties, 
which are universal and model-independent.
The SUSY loop effects on $\bar B \to D \tau \bar\nu_{\tau}$ 
will be discussed elsewhere \cite{MT2}.
}

With these couplings, it turns out that 
the amplitude of charged Higgs exchange 
in $\bar B \to D \tau \bar\nu_{\tau}$ has a term proportional to 
$m_b \tan^2 \beta$. Therefore, the effect of the charged Higgs boson is 
more significant for larger $\tan\beta$. 

Formula of the decay rate is described in Sec.~2.
In Sec.~3, we give hadronic form factors 
including next-to-leading QCD corrections.
In Sec.~4, we show our numerical results. Our conclusion is given in Sec.~5.

\section{Formula of the decay rate}

Using the above Lagrangian in Eq.~(\ref{Lag}) and 
the standard charged current Lagrangian, 
we calculate the amplitudes of charged Higgs exchange and 
$W$ boson exchange in $\bar B \to D \tau \bar\nu_{\tau}$.

The $W$ boson exchange amplitude is given by \cite{Hagiwara}
\bea
   {\cal M}_{s}^{\lambda_\tau}(q^2,x)_W=
   \frac{G_F}{\sqrt{2}}V_{cb} 
   \sum_{\lambda_W}\eta_{\lambda_W} 
   L_{\lambda_W}^{\lambda_\tau}H_{\lambda_W}^{s}\;,\label{amp1}
\ena
where $q^2$ is the invariant mass squared of the leptonic system, and
$x=p_B\cdot p_\tau/m_B^2$.
The $\tau$ helicity and the virtual $W$ helicity are denoted by 
$\lambda_\tau=\pm$ and $\lambda_W=\pm ,\,0,\,s$, and 
the metric factor $\eta_{\lambda_W}$ is given by 
$\eta_{\pm}=\eta_0=-\eta_s=1$.
The hadronic amplitude which describes $\bar B\rightarrow D\,W^*$ 
and the leptonic amplitude which describes 
$W^*\rightarrow\tau\bar\nu$ are given by
\bea
   H_{\lambda_W}^{s}(q^2)&\equiv&
   \epsilon_\mu^*(\lambda_W)\langle D(p_D)|
                    \bar c\gamma^\mu b|\bar B(p_B)\rangle\;,\label{hadamp}\\
   L_{\lambda_W}^{\lambda_\tau}(q^2,x)&\equiv&
   \epsilon_\mu(\lambda_W)\langle\tau(p_\tau,\lambda_\tau)\bar\nu_\tau(p_\nu)|
               \bar \tau\gamma^\mu(1-\gamma_5)\nu_\tau|0\rangle\;,
\ena
where $\epsilon_\mu(\lambda_W)$ is the polarization vector of the virtual
$W$ boson.

The charged Higgs exchange amplitude is given by \cite{Tanaka}
\bea
  {\cal M}_{s}^{\lambda_\tau}(q^2,x)_H=
  \frac{G_F}{\sqrt{2}}V_{cb} L^{\lambda_\tau} 
  \left[X Z^*\frac{m_b m_\tau}{M_{H}^2}
                 H_R^{s}+
                 Y Z^*\frac{m_c m_\tau}{M_{H}^2}
                 H_L^{s}\right]\;.\label{amp2}
\ena
Here, the hadronic and leptonic amplitudes are defined by 
\bea
   H_{R,L}^{s}(q^2)&\equiv&
   \langle D(p_D)|\bar c(1\pm\gamma_5)b|\bar B(p_B)\rangle\;,\label{hadamp2}\\
  L^{\lambda_\tau}(q^2,x)&\equiv&
   \langle\tau(p_\tau,\lambda_\tau)\bar\nu_\tau(p_\nu)|
   \bar \tau (1-\gamma_5)\nu_\tau|0\rangle\;.
\ena
This leptonic amplitude is related to the $W$ exchange amplitude as 
\bea
   L^{\lambda_\tau}=\frac{\sqrt{q^2}}{m_\tau}L^{\lambda_\tau}_s\;.
\ena
Details of the hadronic amplitudes 
for the $W$ exchange and the charged Higgs exchange 
are discussed in the next section.

Using the amplitudes of Eqs.~(\ref{amp1}) and (\ref{amp2}), 
the differential decay rate is given by 
\bea
   \frac{d\Gamma}{dq^2}&=&
   \frac{G_F^2 |V_{cb}|^2 \bar v^4\sqrt{Q_+ Q_-}}{128\pi^3m_B^3}
   \Big[\Big(\frac23 q^2+\frac13 m_{\tau}^2 \Big)(H^s_0)^2 \nonumber\\ 
&&+m_{\tau}^2\Big(
(m_b \tan^2\b +m_c)\frac{\sqrt{q^2}}{M_H^2}H_R^s-H_s^s
\Big)^2 \Big]\;,\label{decayrate}
\ena
where $Q_{\pm}=(m_B\pm m_D)^2-q^2$ and $\bar v=\sqrt{1-m_{\tau}^2/q^2}$.
Note that if $\tan \beta \gtrsim 1$, in which we are interested, 
this decay rate is practically a function of $\tan \beta/M_H$
because the second term in the coefficient of 
$H^s_R$ is negligible for $m_b\tan^2 \beta \gg m_c$.

\section{Hadronic form factors including QCD corrections}

Here, we evaluate the hadronic amplitudes 
in Eq.~(\ref{hadamp}) and Eq.~(\ref{hadamp2}) 
in order to obtain the decay rate numerically.
These amplitudes are given 
in terms of hadronic form factors:
\bea
\langle D(p_D)|\bar c\gamma^\mu b|\bar B(p_B)\rangle 
   &=& \sqrt{m_B m_D}\big[h_{+}(w)(v+v')^{\mu}+h_{-}(w)(v-v')^{\mu}\big]
\;,\\
\langle D(p_D)|\bar c b|\bar B(p_B)\rangle 
   &=& \sqrt{m_B m_D}(1+w)h_{s}(w)
\;,
\ena
where $v=p_B/m_B$, $v'=p_D/m_D$ and 
$w\equiv v\cdot v'=(m_B^2+m_D^2-q^2)/2 m_B m_D$ \cite{HFF}. 

In the heavy quark limit and in the leading logarithmic approximation, 
these form factors $h_{\pm}(w)$ and $h_{s}(w)$ are given as
\bea
 h_{+}(w)=h_{s}(w)=\xi(w)\;,\;\;h_{-}(w)= 0\;,
\ena
where $\xi(w)$ is the universal form factor \cite{Isgur}.

Now, we consider QCD corrections beyond LLA 
and calculate these form factors up to the next-to-leading order. 
Then, these form factors are given as
\bea
h_+(w)&=&\Big\{\hat C_1(w)-\Big(\frac{w+1}{2}\Big)\big(\hat C_2(w)+\hat C_3(w)\big)\Big\}\xi(w)\;,\\
h_-(w)&=&-\Big\{\frac{w+1}{2}\big(\hat C_2(w)-\hat C_3(w)\big)\Big\}\xi(w)\;,\\
h_s(w)&=&\hat C_s(w)\xi(w)\;.
\ena
Explicit formula of coefficients for $W$ exchange, $\hat C_i\;(i=1\sim3)$, 
are given by Neubert \cite{QCDN}.
The coefficient for charged Higgs exchange, $\hat C_s$, is given as
\bea
\hat C_s(w)=A(w)\bar C_s(w)\;,
\ena
where
\bea
A(w)&=&\left(\frac{\a_s(m_c)}{\a_s(m_b)}\right)^{\frac{6}{25}}
[\a_s(m_c)]^{a_L(w)}\;,\label{A}\\
\bar C_s(w)&=&1+\frac{\a_s(m_b)-\a_s(m_c)}{\pi}(\tilde Z+2)
+\frac{\a_s(m_c)}{\pi}\Big[Z(w)+2+\frac32 \big(f(w)-r(w)\big)\Big]\nonumber\\
&&+\frac{2\a_s(\bar m)}{3\pi}g_s(z, w)\;.\label{Cbar}
\ena
$\bar m$ is some average mass of $m_b$ and $m_c$ and 
other functions and constants in Eq.~(\ref{A}) and Eq.~(\ref{Cbar}) 
are given in Ref.\cite{QCDN} and Appendix. 
We have used the $\overline{\rm MS}$ scheme in our calculations.

The form of $\xi(w)$ is strongly constrained by the dispersion relations
as \cite{Caprini}
\bea
 \xi(w)\simeq 1-8\rho_1^2z+(51.\rho_1^2-10.)z^2-(252.\rho_1^2-84.)z^3\;,
\ena
where $z=(\sqrt{w+1}-\sqrt{2})/(\sqrt{w+1}+\sqrt{2})$.
We obtain the slope parameter $\rho_1^2$ as 
\bea
\rho_1^2=1.33\pm 0.22\;,\label{SP}
\ena
from the experimental data of $B\to D^*e\bar\nu_e$ \cite{SPBELLE}.

\section{Numerical results}

We consider the ratio of decay rates, 
\bea
   B=\frac{\Gamma(\bar B\rightarrow D\tau\bar\nu_\tau)}{\Gamma(\bar B\rightarrow D\mu\bar\nu_\mu)_{SM}}\;,\label{Br}
\ena
where the denominator is the decay rate of
$\bar B\rightarrow D\mu\bar\nu_\mu$ in the SM.
Uncertainties due to 
the form factors and other parameters
tend to reduce or vanish by taking the ratio.

\begin{figure}[t]
 \begin{center}
\includegraphics[width=7.8cm]{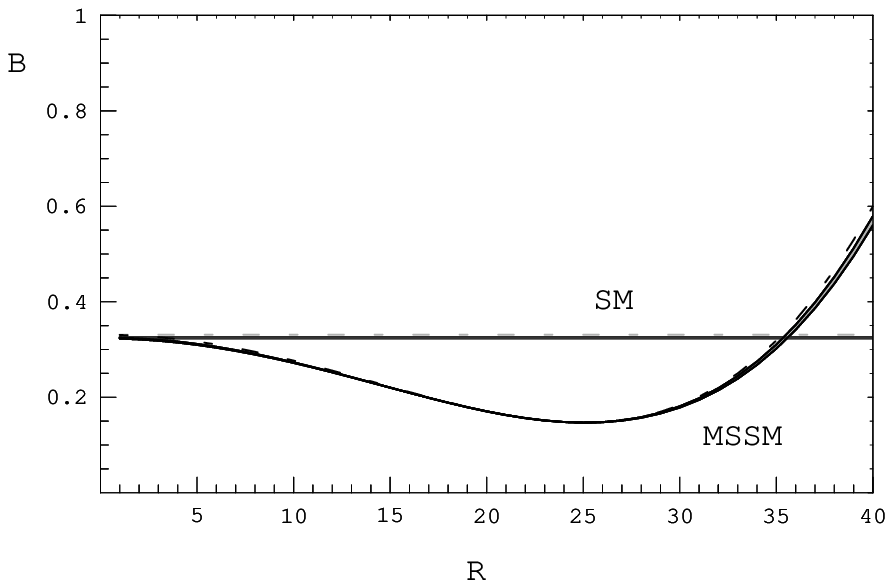} \hspace{0.1cm}
\includegraphics[width=7.8cm]{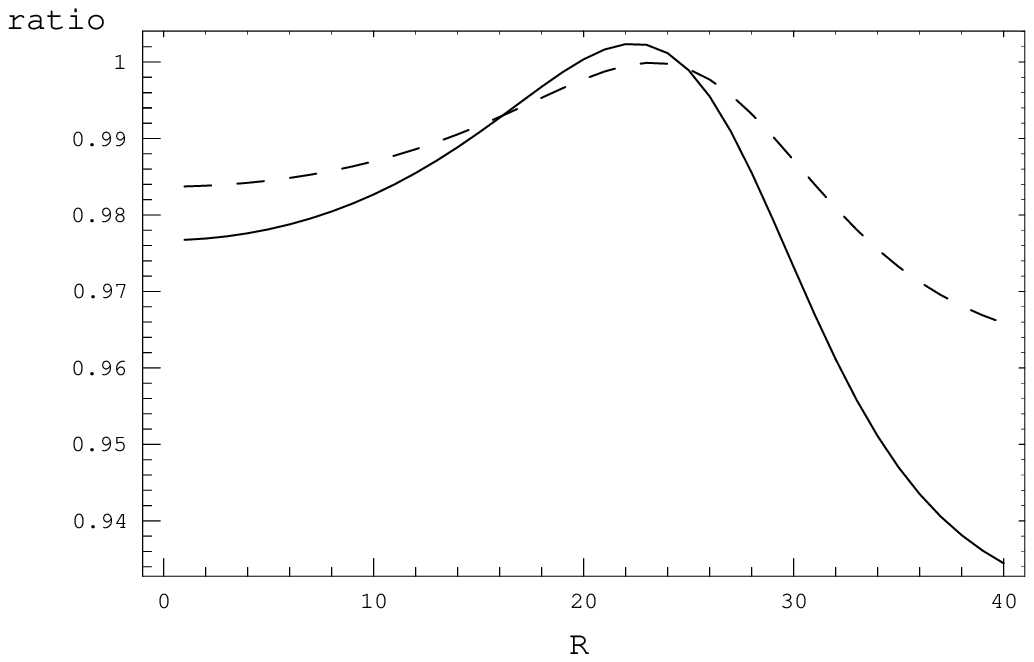}
\caption{(a) The ratio $B$ as a function of $R$ at $\rho_1^2=1.33$ in the MSSM 
  and the SM. The lines with shaded region are obtained by using 
  $\Lambda_{\overline{\rm MS}}=0.15\sim0.25$ GeV and the dashed lines
  show the predictions without QCD corrections.  
  (b) The ratio of $B$ with and without QCD corrections, 
  $B$(with QCD corrections)/$B$(without QCD corrections), as a function of $R$.
  The solid and the dashed lines are 
  the ratios with $\Lambda_{\overline{\rm MS}}=0.25$
  and $0.15$ GeV.}
 \end{center}
\end{figure}

Fig.~1(a) is the plot of our prediction of the ratio in Eq.~(\ref{Br})
as a function of $R$, which is defined by $R\equiv m_W\tan\beta/m_H$. 
Here, we do not show the error in the slope parameter $\rho_1^2$ 
and we take $\rho_1^2=1.33$.
The dashed lines show the MSSM and SM predictions without QCD corrections.
The lines with narrow shaded regions 
show the predictions including QCD corrections
with $\Lambda_{\overline{\rm MS}}$ being varied between 0.15 GeV and 0.25 GeV.

In Fig.~1(b) we show the ratio of $B$ in the MSSM 
with and without QCD corrections, 
$B$(with QCD corrections)/$B$(without QCD corrections), as a function of $R$. 
The solid line is the ratio with $\Lambda_{\overline{\rm MS}}=0.25$ GeV
and the dashed line is that with $\Lambda_{\overline{\rm MS}}=0.15$ GeV.

From Fig.~1, we expect that theoretical uncertainties 
from higher order QCD corrections
are at most a few percents in the ratio of decay rates
and, as seen later, the theoretical uncertainties from QCD corrections are
much smaller than those from the error of $\rho_1^2$.

Fig.~2(a) shows our prediction of the ratio $B$ with QCD corrections
as a function of $R$.
Here, we take the error in the slope parameter into account 
and use $\Lambda_{\overline{\rm MS}}=0.25$ GeV.
The shaded regions show the MSSM and SM predictions with 
the error in the slope parameter $\rho_1^2$ in Eq.~(\ref{SP}).
As mentioned before, from Fig.~2(a) and Fig.~1, 
we see that 
the theoretical uncertainty from the error in the slope parameter 
is dominant over that from QCD corrections.
As seen in Fig.~1(a), when $R$ is about 35, 
the ratio $B$ in the 
MSSM is the same as the one in the SM.

\begin{figure}[t]
 \begin{center}
\includegraphics[width=7.8cm]{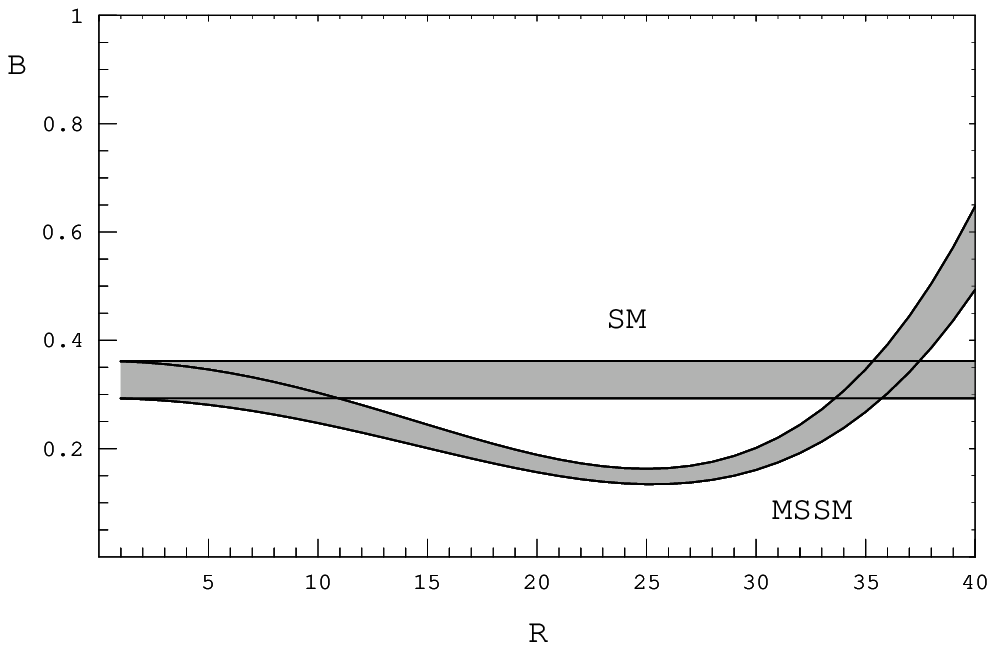} \hspace{0.1cm}
\includegraphics[width=7.8cm]{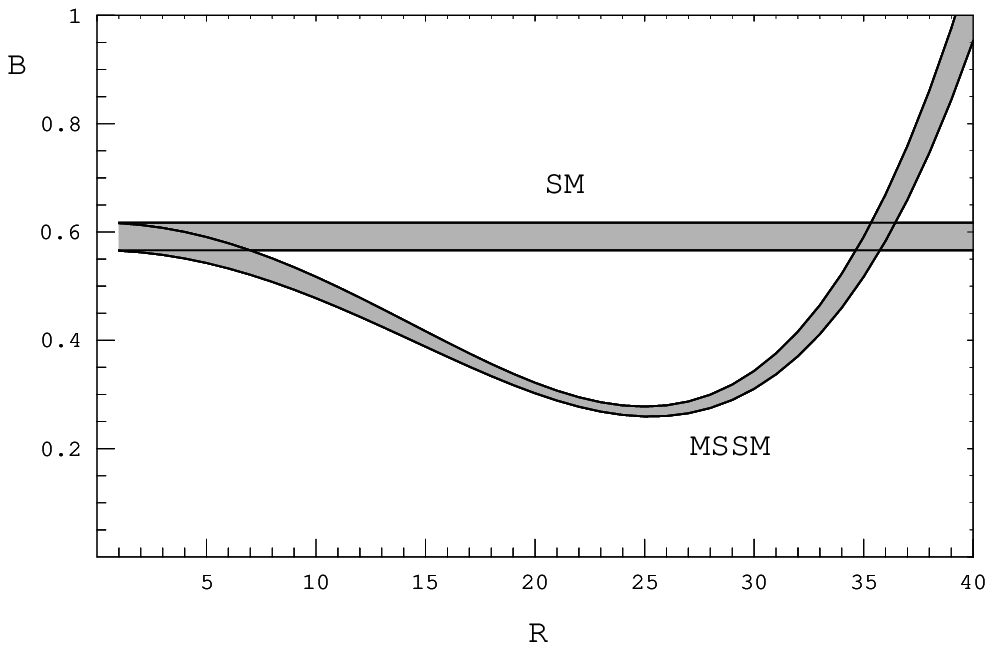}
\caption{The ratios $B$ and $\tilde B$ with QCD corrections 
  as functions of $R$ in the MSSM 
  and the SM. The shaded regions show the predictions with 
  the error in the slope parameter $\rho_1^2$ in Eq.~(\ref{SP}).
  The flat bands show the SM predictions.
  (a) $B$: the decay rate normalized to 
  $\Gamma(\bar B\rightarrow D\mu\bar\nu_\mu)_{SM}$. 
  (b) $\tilde B$: the same as (a) except that 
  the denominator is integrated over 
  $m_{\tau}^2\leq q^2\leq (m_B-m_D)^2$.}
 \end{center}
\end{figure}

In Fig.~2(b), we also show the ratio, 
\bea
   \tilde B=\frac{\Gamma(\bar B\rightarrow D\tau\bar\nu_\tau)}{\tilde\Gamma(\bar B\rightarrow D\mu\bar\nu_\mu)_{SM}}\;,\label{Br2}
\ena
with QCD corrections, similar as Fig.~2(a), but its denominator is 
$\tilde\Gamma(\bar B\rightarrow D\mu\bar\nu_\mu)_{SM}$,
which is integrated over the same $q^2$ region as the $\tau$ mode, 
i.e., $m_{\tau}^2\leq q^2\leq (m_B-m_D)^2$. 
From Fig.~2(b), we observe that the ratio $\tilde B$ has
less theoretical uncertainty and we expect a better sensitivity 
than the ratio $B$ in Fig.~2(a).

\begin{figure}[t]
 \begin{center}
\includegraphics{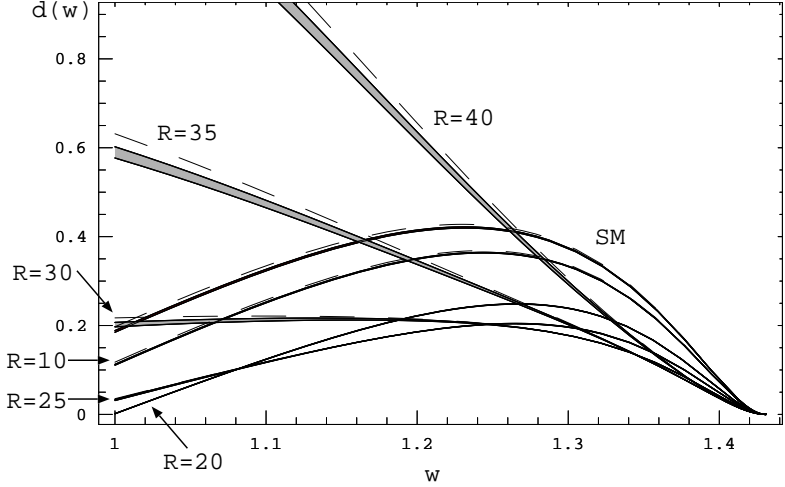}
\caption{The $w$ distribution $d(w)$ 
  in the SM and the MSSM with different values of $R$.
  The lines with shaded regions are given by using 
  $\Lambda_{\overline{\rm MS}}=0.15\sim0.25$ GeV and the dashed lines
  show the prediction without QCD corrections. 
}
 \end{center}
\end{figure}

Now, we consider a decay distribution defined by
\bea
d(w)=(w^2-1)\frac{\Gamma(\bar B\rightarrow D\tau\bar\nu_\tau)/dw}{\Gamma(\bar B\rightarrow D\mu\bar\nu_\mu)_{SM}/dw}\;,
\ena
where $w=(m_B^2+m_D^2-q^2)/2m_Bm_D$.

In Fig.~3, we show $d(w)$ 
in the SM and the MSSM with different values of $R$.
The dashed lines show the MSSM and SM predictions without QCD corrections.
The lines with shaded regions show the predictions with QCD corrections and
the shaded regions are given by using 
$\Lambda_{\overline{\rm MS}}=0.15\sim0.25$ GeV.
The theoretical uncertainty from 
the error in the slope parameter is canceled out in this quantity.
Thus, QCD corrections become 
dominant uncertainties in the theoretical calculation.
From Fig.~2, the ratio $B$ in the 
MSSM becomes the same as the one in the SM when $R\sim35$.
But, in Fig.~3, we find that 
the behavior of $d(w)$ in the MSSM with $R=35$ is considerably different from 
that in the SM. Therefore, we can distinguish the SM from the MSSM 
by investigating $w$ distribution even if $R\sim35$.

\section{Conclusion}
As seen in our numerical results, the branching ratio of
$\bar B\rightarrow D\tau\bar\nu_\tau$
is a sensitive probe of the MSSM-like Higgs sector. 
So, if $\bar B\rightarrow D\tau\bar\nu_\tau$ is observed 
at a B factory experiment, a significant region 
of the parameter space of the MSSM Higgs sector will be proved.
This is complementary at the Higgs search at LHC \cite{LHC}.

In the branching ratio, the theoretical uncertainty from QCD correction
is much smaller than that from the error in the slope parameter $\rho_1^2$.
However, in $w$ distribution, the theoretical uncertainty 
from the error in the slope parameter $\rho_1^2$ is canceled out and, therefore,
it is important to consider QCD corrections.

In this work, we have not taken $1/m$ corrections into account.
They may be as significant as QCD corrections in the $w$ distribution.
However we expect that their effects are smaller than the uncertainty
from the error of $\rho_1^2$ in the branching ratio \cite{MT2}.

Finally, the $\tau$ polarization 
in $\bar B\rightarrow D\tau\bar\nu_\tau$ is also expected 
to be a good probe of charged Higgs boson.
The theoretical uncertainty from the error in the 
slope parameter becomes very small in the $\tau$ polarization \cite{Tanaka}.
QCD and $1/m$ corrections to this quantity will be addressed elsewhere.

\setcounter{section}{1}
\renewcommand{\thesection}{\Alph{section}}
\renewcommand{\theequation}{\thesection .\arabic{equation}}
\setcounter{equation}{0}

\section*{Appendix}
Functions in Eq.~(\ref{A}) and Eq.~(\ref{Cbar}) are given by
\bea
r(w)&=&\frac{1}{\sqrt{w^2-1}}\ln(w+\sqrt{w^2-1})\;,\\
f(w)&=&wr(w)-2-\frac{w}{2\sqrt{w^2-1}}\Big[L_2(1-w_-^2)-L_2(1-w_+^2)\Big]\;,\\
g_s(z, w)&=&\frac{w}{\sqrt{w^2-1}}\Big[L_2(1-zw_-)-L_2(1-zw_+)\Big]\nonumber\\
&&-\frac{z}{1-2wz+z^2}\Big[(w^2-1)r(w)+(w-z)\ln z\Big]\;,\\
a_L(w)&=&\frac{8}{27}\Big[wr(w)-1\Big]\;,\\
Z(w)&=&-\frac49 \Big[\frac{25}{54}+\frac{\pi^2}{12}+\frac{5}{9}+f(w)\Big]
\Big[wr(w)-1\Big]-\frac89 I(w)\;,\\
\tilde Z&=&-\frac{7}{225}\pi^2-\frac{9403}{7500}\;,
\ena
where $z=m_c/m_b$, $w_{\pm}=w\pm\sqrt{w^2-1}$,
\bea
L_2(x)=-\int_0^x dt\frac{\ln(1-t)}{t}\;,
\ena
is the dilogarithm and,
\bea
I(w)&=&\int^\varphi_0 d\psi\Big[\psi\coth\psi-1\Big]\nonumber\\
&&\times \Big\{\psi\coth^2\varphi
+\frac{\sinh\varphi\cosh\varphi}{\sinh^2\varphi-\sinh^2\psi}
\ln\frac{\sinh\varphi}{\sinh\psi}\Big\}\;,
\ena
in terms of the hyperbolic angle $\varphi$ defined by $w=\cosh\varphi$.


\begin{thebibliography}{99}

\bibitem{MSSM}
For a review, see, e.g., 
H.~E.~Haber and G.~L.~Kane, Phys. Rep. 117 (1985) 75.

\bibitem{MT}
T.~Miura and M.~Tanaka, hep-ph 0109244.

\bibitem{Tanaka}
M.~Tanaka, Z. Phys. C67 (1995) 321.

\bibitem{Soni} K.~Kiers and A.~Soni, \PR{D56} (1997) 5786.

\bibitem{CKM}
M.~Kobayashi and T.~Maskawa, 
\PTP{49} (1973) 652.

\bibitem{HHG}
J.~F.~Gunion, H.~E.~Haber, G.~L.~Kane and S.~Dawson, 
{\em The Higgs Hunter's Guide}
(Addison-Wesley Publishing Company, 1990).


\bibitem{Wagner}
M.~Carena, D.~Garcia, U.~Nierste and C.~E.~Wagner,
\NP{B577} (2000) 88.

\bibitem{Sola}
J.~A.~Coarasa, R.~A.~Jimenez and J.~Sola,
\PL{B406} (1997) 337.


\bibitem{SUSY-QCD}
L.~J.~Hall, R.~Rattazzi and U.~Sarid, 
\PR{D50} (1994) 7048;\\
R.~Hempfling, 
\PR{D49} (1994) 6168.


\bibitem{MT2}
T.~Miura and M.~Tanaka, in preparation.

\bibitem{Hagiwara}
K.~Hagiwara, A.~D.~Martin and M.~F.~Wade, 
\NP{B327} (1989) 569;\\
K.~Hagiwara, A.~D.~Martin and M.~F.~Wade, Z. Phys. C46 (1990) 299.

\bibitem{HFF}
M.~Neubert, \PL{B264} (1991) 455.

\bibitem{Isgur}
N.~Isgur and M.~B.~Wise, \PL{B232} (1989) 113; \PL{B237} (1990) 527.

\bibitem{QCDN}
M.~Neubert, \PR{D46} (1992) 2212.

\bibitem{Caprini} 
I.~Caprini, L.~Lellouch and M.~Neubert, 
\NP{B530} (1998) 153.

\bibitem{SPBELLE} K.~Abe {\it et.al.}, BELLE-CONF-0121 (2001).

\bibitem{LHC} F.~Gianotti, talk presented at LHCC, 5 July, 2000,\\
http://gianotti.home.cern.ch/gianotti/phys\_info.html .

\end{thebibliography}
\end{document}